\documentclass[%
 aip,
 amsmath,amssymb,
 reprint,%
]{revtex4-1}
\usepackage{graphicx}
\usepackage{dcolumn}
\usepackage{bm}

\usepackage[utf8]{inputenc}
\usepackage[T1]{fontenc}
\usepackage{mathptmx}
\usepackage{etoolbox}
\usepackage{physics}
\usepackage[shortlabels]{enumitem}

\makeatletter
\def\@email#1#2{%
 \endgroup
 \patchcmd{\titleblock@produce}
  {\frontmatter@RRAPformat}
  {\frontmatter@RRAPformat{\produce@RRAP{*#1\href{mailto:#2}{#2}}}\frontmatter@RRAPformat}
  {}{}
}%
\makeatother

\begin{document}
\preprint{AIP/123-QED}
\title{Frequency-resolved Transient Absorption Spectroscopy for High Pressure System}
\author{Zi-Qian Cheng}
\affiliation{Graduate School of China Academy of Engineering Physics, Beijing 100193, China}
\author{Xiao-Shuang Yin}
\affiliation{Center for High Pressure Science and Technology Advanced Research, Beijing 100193, China}
\author{Liu-Xiang Yang}
\email{liuxiang.yang@hpstar.ac.cn}
\affiliation{Center for High Pressure Science and Technology Advanced Research, Beijing 100193, China}
\author{Hui Dong}
\email{hdong@gscaep.ac.cn}
\affiliation{Graduate School of China Academy of Engineering Physics, Beijing 100193, China}

\begin{abstract}
Dynamics of materials under high-pressure conditions has been an important
focus of materials science, especially in the timescale of pico- and
femto-second of electronic and vibrational motion, which is typically
probed by ultrafast laser pulses. To probe such dynamics, it requires
an integration of high-pressure devices with the ultrafast laser system.
In this work, we construct a frequency-resolved high-pressure transient
absorption spectroscopy system based on a diamond anvil cell (DAC)
with transmissive detection. In this setup, we use the narrowband
laser as the pump beam and the supercontinuum white light as the probe
beam. To effectively eliminate the scattering noise from the pump
light, we design a double-chopper operating mode, which allows us
to obtain signals in the complete frequency domain including the overlap
region with the pump pulse. And we test system with Rhodamine B solution
with the probe wavelength range of 450-750 nm and the 550nm pump,
and observe that the intensity of the signal peak corresponding to
the monomer at 560 nm continuously decreased relative to the signal
peak corresponding to the dimer at 530 nm. This indicates that the
portion of Rhodamine B molecules in the dimer form increases under
increasing pressure. Additionally, we find two dynamic components
of the signal peaks for both monomer and dimer, and the short-lifetime
component increases as the pressure is increased, and the long-lifetime
component decreases.
\end{abstract}
\maketitle

\section{Introduction}

With the advent of femtosecond pulse laser technology, transient absorption
(TA) spectroscopy has developed rapidly and has been widely used in
molecular biology, chemistry, and condensed matter physics. Its femtosecond
time resolution enables to explore the charge and energy transfer
dynamics in researches of photosynthetic proteins\citep{Berera2009,Croce2001,Kaucikas2017,Zamzam2020}, organic molecules\citep{Hunt2019}, dyes\citep{Jiang2013,Dobryakov2005,Maly2017}, semiconductor devices\citep{Ohkita2011,Ravensbergen2014,Knowles2018},
and perovskites\citep{Yarita2017,Wang2014,Serpetzoglou2017}. Currently,
most of the applications remain confined to normal pressure environments.
To extend spectroscopic methods into the high-pressure researches,
it requires an integration of high-pressure apparatus with the spectroscopic
detections. One of the possible apparatus is the diamond anvil cell
(DAC) \citep{Bridgman1950,Bassett2009}, which has played a crucial
role in the field of high-pressure physics as a safe, low-cost, easily
operable experimental device capable of applying pressures over 100
GPa\citep{Jayaraman1983}. It combines high pressure with high temperature,
strong magnetic fields, and optical excitation, assisting researchers
in exploring the physical properties of materials under various conditions\citep{Bassett2009}.
The transparent windows on the both sides of DAC offer a possibility
to integrate the laser pulse probe. However, due to the effects of
light absorption and scattering by the diamonds within the DAC, the
application of femtosecond pulsed for measurements of samples inside
the DAC still faces many limitations.

There are already several successful experimental attempts to combine
high pressure with spectroscopy\citep{Zhou2020}, including the detection
of the absorption and photoluminescence spectra under pressure gradients\citep{Liu2023}
with the spectrum filter \citep{Wu2021}. Due to the limitations of
aperture and sample size in DAC, co-linearly arranging of the pump
and probe beams of different wavelengths is typically used along with
bandpass filters to block scattered noise. In such a setup, it is
hard to have the frequency resolution \citep{Li2020}.

To obtain dynamic information, it typically requires the spectrum
overlap between the pump and probe pulse under resonant pump-probe
conditions, where the scattering from the pump pulse results in significant
amount of noise. To overcome such problem, we design a TA system with
non-collinear geometry so that only a small portion of the pump light
is scattered on the diamond into the probe light. By adding choppers
in the both pump and probe beams, we further remove the scattering
noise from the pump and obtain the spectrum of the pump scattering
noise in real time and subtract it, such that a complete spectrum
is obtained even when the spectrum of the pump beam is overlapped
with the probe beam. With this setup, we use Rhodamine B in water
as the test sample and observe clear spectral changes in the pressure
range of 0.11-0.98 GPa. We find two ground-state-bleaching peaks corresponding
to the Rhodamine B monomers and dimers, and their intensity changes
with the increasing pressure. Such change characterize the process
in which the Rhodamine B molecules in water gradually transform from
monomers to dimers.

\begin{figure}[tbph]
\includegraphics{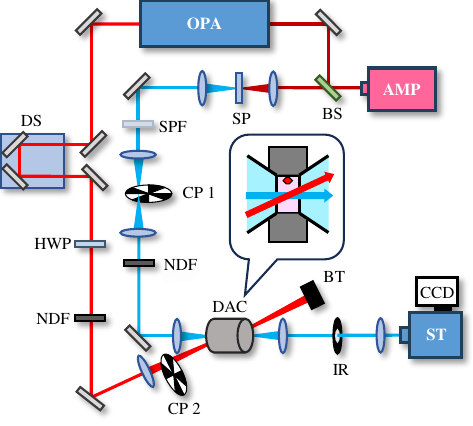}
\caption{The experimental setup of frequency-resolved high-pressure transient
absorption spectroscopy. AMP: Ti:Sapphire amplifier laser system;
OPA: optical parametric amplifier; BS: beam splitter; SP: sapphire
window; SPF: short-pass filter; DS: delay stage; CP: chopper; HWP:
half-wave plate; NDF: neutral density filter; DAC: diamond anvil cell;
BT: beam trap; IR: iris; ST: spectrometer; CCD: charge coupled device
camera. Magnified view: the sample cell structure, where the light
blue area, dark gray area, pink area, and red hexagon represent the
diamond, rhenium plate, Rhodamine B solution, and ruby, respectively.
In the entire diagram, the dark red line, blue line, and red line
represent the 800 nm fundamental pulse, the probe pulse, and the pump
pulse, respectively.}
\label{fig:The-experimental-setup}
\end{figure}

\section{Experimental set-up}

In Fig.\ref{fig:The-experimental-setup}, we present the current setup
of the frequency-resolved high-pressure transient absorption spectroscopy
system. The light source for TA spectra is an 800-nm Ti:sapphire amplified
pulsed laser which delivers pulses with the width of 35fs at a 10-kHz
repetition rate. The output light is split into two parts: one beam
(red lines) is used to generate a 550 nm pump pulse through an optical
parametric amplifier, and the other beam (blue lines) is used to generate
supercontinuum white light via a 3 mm thick sapphire window as the
probe pulse\citep{Dubietis2019,Dharmadhikari2005}. The pump pulse
is sent through an electronically controlled mechanical delay line
to adjust the optical path length, thereby controlling the time delay
between the pump and probe pulses reaching the sample. The probe light
is incident vertically on the sample in the DAC with pulse energies
of 0.58nJ, while the pump light is incident on the sample at an angle
of approximately 15° to the normal with pulse energies of 9nJ. The
polarization angle between the two beams is set to the magic angle
to avoid noise caused by anisotropy of dipoles\citep{Schott2014}.
We use lenses with focal lengths of 150 mm and 200 mm to focus the
probe and pump pulses, respectively. This setup ensures that the two
pulses can completely incident on the sample and the pump beam fully
covers the probe beam at the sample location, maximizing the signal
intensity. After passing through the sample, the probe pulse is led
into the spectrometer and detected by a charge coupled device (CCD)
camera.
\begin{figure}[tbph]
\includegraphics{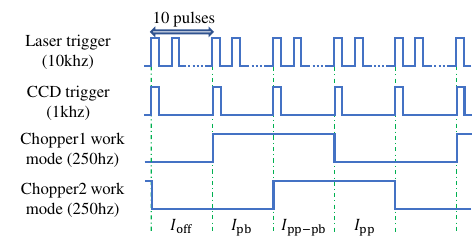}
\caption{Timing logic for double-chopper mode in our transient absorption system.}
\label{fig:Timing-logic-for}
\end{figure}

The DAC with a culet diameter of 400 {\textmu}m is used
for the pressure loading. A Rhenium (Re) plate with thickness of 250
micrometers is used as the gasket. After pre-indentation to 70 {\textmu}m,
a 250 {\textmu}m diameter hole is drilled into the gasket
by laser to create a specimen chamber. The sample is placed inside
the chamber, which is sandwiched by the diamonds. A ruby is placed
in the chamber for the pressure calibration, and is positioned close
to the edge of the chamber to avoid being struck by any pump or probe
pulses to generate artifacts in the signal. Between each set of TA
spectra experiments, the DAC is removed from the optical path and
pressurized. And the wavelength of the ruby fluorescence characteristic
peak is measured to calibrate the pressure within the chamber for
different experimental sets. To keep the sample in the liquid phase,
we set the maximum pressure within the chamber as 1.0 GPa, with increments
of 0.05 GPa for each experiments. In the liquid phase, the pressure
distribution is uniform throughout the sample, and thus the position
of the ruby does not affect the accuracy of the pressure measurement.

In the current setup, we utilize a double-chopper setup \citep{Brixner2004,Yue2022}
to eliminate the intense scattering noise generated when the pump
pulse passes through the diamond in the DAC and other time-varying
system noises. We balance the CCD's acquisition area size and sampling
frequency, setting the exposure frequency to 1 kHz, which corresponds
to capturing 10 pulses with each exposure. Two choppers, each with
a frequency of 250 Hz, are positioned in the pump and probe beam paths,
respectively. Due to the beam diameter of the supercontinuum white
light being approximately 10 mm, we use a pair of achromatic lenses
to focus and collimate the probe pulse before and after the chopper.
This arrangement ensures that every 10 pulses are either completely
blocked or fully transmitted by the chopper blades, with the chopper
positioned at the focal points of the two lenses.

The laser trigger signal is processed into synchronization control
signals for the CCD and choppers using a digital circuit board. By
setting the relative phases, we set up the timing logic of our system
as illustrated in Fig.\ref{fig:Timing-logic-for}. The laser trigger
signal with 10kHz is processed to generate the CCD signal at $1$kHz
to trigger the camera. The two choppers are triggered by the signal
at $250$Hz with different phase shift to form four different exposure
combinations. Each acquisition cycle consists of these four exposures,
pump-on and probe-on ($I_{\mathrm{pp-pb}}$), pump-off and probe-on
($I_{\mathrm{pb}}$), pump-on and probe-off ($I_{\mathrm{pp}}$),
and pump-off and probe-off ($I_{\mathrm{off}}$). And the final TA
signal is obtained as
\begin{equation}
\Delta T/T  =\frac{(I_{\mathrm{pp-pb}}-I_{\mathrm{pb}})-(I_{\mathrm{pp}}-I_{\mathrm{off}})}{I_{\mathrm{pb}}}.
\end{equation}
Scattering noise from the pump pulse is included in $I_{\mathrm{pp-pb}}$
and $I_{\mathrm{pp}}$, and in turn eliminated from the final results. 

\begin{figure}[tbph]
\includegraphics{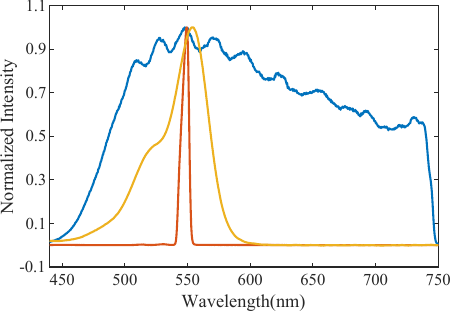}
\caption{The spectrum of the pump light (red line), the spectrum of the probe light (blue line), and the absorption spectrum of Rhodamine B under the ambient
pressure (yellow line).}
\label{fig:The-pump-light}
\end{figure}

\section{Results and discussions}

With the current setup, we perform TA experiments on the sample with
Rhodamine B in the water. A series of TA spectra are measured under
different pressures ranging from 0.11 GPa to 0.98 GPa. The pump is
a femtosecond laser pulse with central wavelength $550\mathrm{nm}$
and its spectral is shown as the red line in Fig.\ref{fig:The-pump-light}.
The supercontinuum white light with wavelength from 450 nm to 750
nm is used as the probe pulse, whose spectral is illustrated as the
blue line in Fig.\ref{fig:The-pump-light}. The absorption spectrum
of Rhodamine B under the the ambient pressure is measured and shown
in Fig.\ref{fig:The-pump-light} as the yellow line. The central wavelength
of the pump pulse is chosen to ensure the effective excitation of
the Rhodamine B molecules.

\begin{figure}
\includegraphics{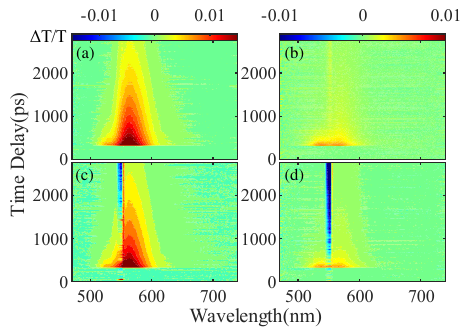}
\caption{The TA spectra of the Rhodamine B in water. (a) and (b) TA results
at 0.11 GPa and 0.85 GPa under double-chopper mode, respectively.
(c) and (d) TA results at 0.11 GPa and 0.85 GPa under the single-chopper
mode, respectively.
\label{fig:The-TA-spectra}}
\end{figure}

We show the obtained TA spectra for two typical pressures 0.11 GPa
and 0.85 GPa in Fig.\ref{fig:The-TA-spectra}(a) and (b) respectively
with the current setup using double choppers. The data have been processed
using the moving window smoothing technique along the wavelength axis
to eliminate the oscillation interference caused by multiple reflections
of the probe light on the diamond surface. The range of smoothing
window is set to 5 adjacent pixels (2.5 nm for the wavelength range),
ensuring effective noise filtering without losing signal peaks. Both
TA spectra show the similar ground-state bleaching peaks at 560nm.
Comparing the signal dynamics in Fig.\ref{fig:The-TA-spectra}(a)
and (b) along the time delay axis, we observe that the relaxation
is significantly faster at 0.85 GPa than that at 0.11 GPa. To show
the effectiveness of our double choppper scheme, we also present the
TA signals obtained using the traditional single-chopper method in
Fig.\ref{fig:The-TA-spectra}(c) and \ref{fig:The-TA-spectra}(d),
where strong pump scattering noise is observed near 550 nm as the
blue color data. In comparison, the scattering noise has been almost
completely eliminated in Fig.\ref{fig:The-TA-spectra}(a) and (b),
with its intensity reduce to less than one-tenth of the maximum signal
peak. The significant reduction of the noise related to the pump light
scattering is one of advantages of the current setup, which allows
the measurement of the full spectrum as presented in Fig.\ref{fig:The-TA-spectra}(a)
and (b). The results in Fig.\ref{fig:The-TA-spectra}(a) are consistent
with the previously reported TA spectra of Rhodamine B in water under
ambient pressure \citep{Maly2017}.

\begin{figure}
\includegraphics{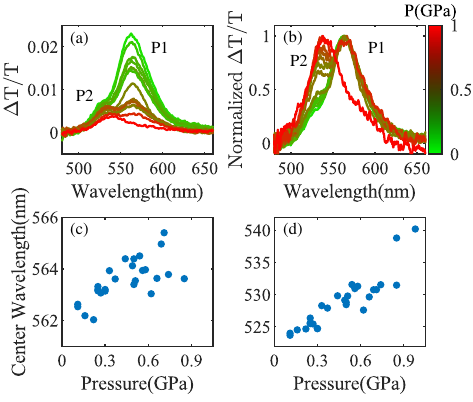}
\caption{The raw data (a) and normalized results (b) of TA spectra at the time
delay of 50 ps under different pressure ranging from 0.11 GPa to 0.98
GPa. The pressure increases from the lowest (green line) to the highest
(red line). And the changes of the central wavelengths of the peak
P1 (c) and P2 (d) with increasing pressure.
\label{fig:cutandcenter}}
\end{figure}

In Fig.\ref{fig:cutandcenter}, we show slice of the spectra at the
time delay of 50 ps, and present the raw data in Fig.\ref{fig:cutandcenter}
(a) and the normalized results in Fig.\ref{fig:cutandcenter} (b)
for different pressures. The spectra show two distinct ground-state
bleaching signal peaks at 560 nm (peak P1) and 531 nm (peak P2). Here
the peak P1 at 560 nm corresponds to the main absorption peak of Rhodamine
B monomers and the peak P2 at 530 nm corresponds to the absorption
of dimers\citep{Fujii1995}. The raw data shows that both intensities
of the peak P1 and P2 decrease with the increase of pressure. The
decay rate for the intensity of the peak P1 is larger than that of
the peak P2, and in turn the intensity of the peak P2 is higher than
that of peak P1 with the pressure higher than 0.74 GPa. The changes
in the relative intensity of the signal peaks are attributed to the
pressure-induced conversion of Rhodamine B molecules from monomers
to dimers within the sample, which is consistent with previous experiments\citep{Roberts1985}.

To track the impact of the pressure on the transient absorption, we
select the localized signal within the range from 558 nm to 570 nm
for Gaussian fitting to obtain the profile and center wavelength of
peak P1. Then, we subtract the fitted Gaussian peak of P1 from the
520 - 548 nm signal and perform Gaussian fitting again to obtain the
profile and center wavelength of P2. The change of the center wavelengths
of P1 and P2 with pressure is shown in Fig.\ref{fig:cutandcenter}(c)
and (d). When the pressure exceeds 0.85 GPa, the intensity of P1 becomes
so weak that it is completely overshadowed by P2. Therefore, the center
wavelength of P1 is not shown at this pressure, and the center wavelength
of P2 is directly obtained from the fitting of the original signal.
As shown in Fig.\ref{fig:cutandcenter}(c) and (d), with the pressure
increase, the center of peak P1 redshifts from 562.0 nm to 565.4 nm
, while the center of P2 redshifts from 523.7 nm to 540.2 nm. Such
shift of the peak position may be caused by reduction of the reorganization
energy of the transition due to the freezing of the freedom of the
solution as the environment. With the pressure higher than 0.98 Gpa,
the solution is turned into the solid phase.

\begin{figure}
\includegraphics{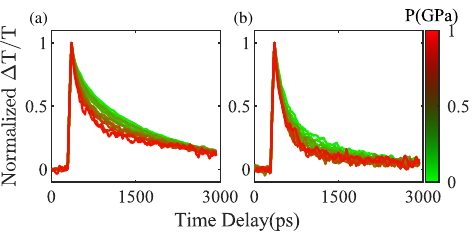}
\caption{The relaxation of the peak P1 (a) and peak P2 (b) with time delay.
The data in the figure are normalized by the intensity at the time
delay when the pump and probe pulses are completely overlapped. The
pressure increases from the lowest (green line) to the highest (red
line).
\label{fig:The-relaxation-of-signal}}
\end{figure}

\begin{figure}
\includegraphics{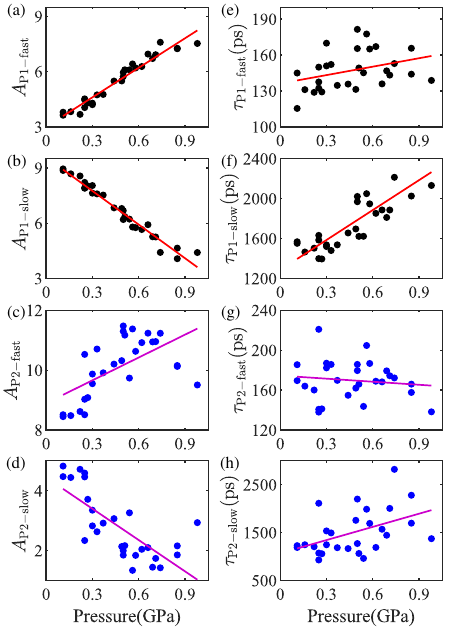}
\caption{The amplitudes and lifetimes of the peak P1 and peak P2 as a function
of pressure. (a), (b), (e) and (f) Pressure-induced changes in amplitude
and lifetime of the fast and slow component of the peak P1. The red
lines show the trend of changes. (c), (d), (g) and (h) Pressure-induced
changes in amplitude and lifetime of the fast and slow component of
the peak P2. The purple lines represent the trend of changes \label{fig:The-pressure-dependence}}
\end{figure}

To extract the impact of the pressure on the dynamics, we plot the
intensities of signals within a 5 nm wavelength range around peak
P1 and peak P2 as a function of the delay time, shown in Fig.\ref{fig:The-relaxation-of-signal}.
These curves show a clear change of the dynamics of the signal with
the increase of the pressure. The red curve (high pressure) exhibits
a significantly deeper concavity compared to the green curve (low
pressure). To quantitatively characterize these signals, we decompose
the intensity changes of the peak P1 and P2 into two kinetic components
and display the changes of their amplitude and lifetimes with pressure
in Fig.\ref{fig:The-pressure-dependence} by fitting the dynamics
with the formula $\Delta T/T=A_{\mathrm{Pi-slow}}\exp\left(-t/\tau_{\mathrm{Pi-slow}}\right)+A_{\mathrm{Pi-fast}}\exp\left(-t/\tau_{\mathrm{Pi-fast}}\right)$
$\left(i=1,2\right)$. The trends of the fitting parameters for peak
P1 and peak P2 are represented by the red and purple lines, respectively. 

As shown in Fig.\ref{fig:The-pressure-dependence}(a), (b), (c), and
(d), the amplitudes of the fast components of both peaks increase
rapidly with increasing pressure, while the amplitudes of the slow
components decrease. Such trend of shifting from the slow component
to the fast component is consistent with the data within 1000 ps of
the time delay in Fig.\ref{fig:The-relaxation-of-signal}.

In terms of lifetime, Fig.\ref{fig:The-pressure-dependence}(e) and
(g) show almost no changes in the lifetime of the fast component of
the single peaks with the changing pressure. Such an observation indicates
that the kinetic mechanism of the rapid decay is not affected by the
pressure. However, both the lifetime of the slow components in Fig.\ref{fig:The-pressure-dependence}(f)
and (h) increase rapidly as the pressure rises. This lifetime is close
to the fluorescence lifetime of Rhodamine B in water\textbf{ }under
ambient pressure\citep{Kristoffersen2014}.

We attribute the fast component to the inter-molecule interaction,
whose amplitude is strongly affected by the pressure due to the change
in molecular distance between the molecules. As the molecular distance
decreases, more excited-state energy is transferred to solvent molecules
through inter-molecular interactions instead of intra-molecular relaxation.
However, the dynamics of this channel remain unchanged, as indicated
by the consistency of its lifetime across different pressures.

For the slow component, we believe it represents intra-molecular relaxation,
which is determined by the molecular structure of Rhodamine B. The
pressure-induced transition of rhodamine molecules from monomers to
dimers leads to an increase of the intramolecular relaxation lifetime
\citep{Monte1998}, which is consistent with the results shown in
Fig.\ref{fig:The-pressure-dependence}(f) and (h). The amplitude of
this relaxation process decreases as the contribution of inter-molecular
interactions increases. At the same time, with the increase of the
pressure, the solution transitions from the liquid phase to the solid
phase, where some of the molecular degrees of the freedom is frozen
and in turn result in longer lifetime. Indeed, we observe a substantial
decrease in their absorption of the pump light when the pressure exceeds
0.98 GPa, due to the impending phase transition of the sample from
the liquid to the solid phase.

\section{Conclusions}

In conclusion, we construct a frequency-resolved high-pressure transient
absorption spectroscopy system and conduct tests using Rhodamine B
in water as the sample. This system uses the narrowband pulse as the
pump beam and the supercontinuum white light as the probe beam, allowing
for the acquisition of complete spectral signals in the 450-750 nm
range, as well as performing dynamic analysis. By utilizing a double-chopper
setup, the system effectively mitigates the impact of pump light scattering
on the measurement results, allowing the measurement of the full TA
spectrum when the pump and probe pulse have spectral overlap. This
technique significantly extends the system's usability, making it
possible to investigate the dynamic processes following resonant excitation
under high-pressure conditions. The system is potentially applicable
to samples that allow for stable light transmission, such as liquid
samples and thin films.

\section*{acknowledgments}

This work is supported by the Innovation Program for Quantum Science
and Technology (Grant No. 2023ZD0300700), and the National Natural
Science Foundation of China (Grant Nos. U2230203, U2330401, and 12088101).

\section*{AUTHOR DECLARATIONS}

\subsection*{Conflict of Interest}

The authors have no conflicts to disclose.

\subsection*{Author Contributions}

\noindent{\bf Zi-Qian Cheng:} Design and construction of the experimental
system; Data acquisition (equal); Formal analysis (equal); Manuscript
writing - review and editing (equal). {\bf Xiao-Shuang Yin: }Data
acquisition (equal); Manuscript review and editing (supporting). {\bf Liu-Xiang
Yang: }Conceptualization (equal); Formal analysis (equal); Manuscript
review and editing (supporting). {\bf Hui Dong: }Conceptualization
(equal); Formal analysis (equal); Funding acquisition (lead); Project
administration (lead); Writing - review and editing (lead).

\section*{DATA AVAILABILITY}

The data that support the findings of this study are available from
the corresponding author upon reasonable request.

\bibliographystyle{apsrev4-1}
\bibliography{FSHPTA}

\begin{thebibliography}{30}%
\makeatletter
\providecommand \@ifxundefined [1]{%
 \@ifx{#1\undefined}
}%
\providecommand \@ifnum [1]{%
 \ifnum #1\expandafter \@firstoftwo
 \else \expandafter \@secondoftwo
 \fi
}%
\providecommand \@ifx [1]{%
 \ifx #1\expandafter \@firstoftwo
 \else \expandafter \@secondoftwo
 \fi
}%
\providecommand \natexlab [1]{#1}%
\providecommand \enquote  [1]{``#1''}%
\providecommand \bibnamefont  [1]{#1}%
\providecommand \bibfnamefont [1]{#1}%
\providecommand \citenamefont [1]{#1}%
\providecommand \href@noop [0]{\@secondoftwo}%
\providecommand \href [0]{\begingroup \@sanitize@url \@href}%
\providecommand \@href[1]{\@@startlink{#1}\@@href}%
\providecommand \@@href[1]{\endgroup#1\@@endlink}%
\providecommand \@sanitize@url [0]{\catcode `\\12\catcode `\$12\catcode `\&12\catcode `\#12\catcode `\^12\catcode `\_12\catcode `\%12\relax}%
\providecommand \@@startlink[1]{}%
\providecommand \@@endlink[0]{}%
\providecommand \url  [0]{\begingroup\@sanitize@url \@url }%
\providecommand \@url [1]{\endgroup\@href {#1}{\urlprefix }}%
\providecommand \urlprefix  [0]{URL }%
\providecommand \Eprint [0]{\href }%
\providecommand \doibase [0]{http://dx.doi.org/}%
\providecommand \selectlanguage [0]{\@gobble}%
\providecommand \bibinfo  [0]{\@secondoftwo}%
\providecommand \bibfield  [0]{\@secondoftwo}%
\providecommand \translation [1]{[#1]}%
\providecommand \BibitemOpen [0]{}%
\providecommand \bibitemStop [0]{}%
\providecommand \bibitemNoStop [0]{.\EOS\space}%
\providecommand \EOS [0]{\spacefactor3000\relax}%
\providecommand \BibitemShut  [1]{\csname bibitem#1\endcsname}%
\let\auto@bib@innerbib\@empty
\bibitem [{\citenamefont {Berera}\ \emph {et~al.}(2009)\citenamefont {Berera}, \citenamefont {van Grondelle},\ and\ \citenamefont {Kennis}}]{Berera2009}%
  \BibitemOpen
  \bibfield  {author} {\bibinfo {author} {\bibfnamefont {R.}~\bibnamefont {Berera}}, \bibinfo {author} {\bibfnamefont {R.}~\bibnamefont {van Grondelle}}, \ and\ \bibinfo {author} {\bibfnamefont {J.~T.~M.}\ \bibnamefont {Kennis}},\ }\href {\doibase 10.1007/s11120-009-9454-y} {\bibfield  {journal} {\bibinfo  {journal} {Photosynth. Res.}\ }\textbf {\bibinfo {volume} {101}},\ \bibinfo {pages} {105} (\bibinfo {year} {2009})}\BibitemShut {NoStop}%
\bibitem [{\citenamefont {Croce}\ \emph {et~al.}(2001)\citenamefont {Croce}, \citenamefont {Muller}, \citenamefont {Bassi},\ and\ \citenamefont {Holzwarth}}]{Croce2001}%
  \BibitemOpen
  \bibfield  {author} {\bibinfo {author} {\bibfnamefont {R.}~\bibnamefont {Croce}}, \bibinfo {author} {\bibfnamefont {M.~G.}\ \bibnamefont {Muller}}, \bibinfo {author} {\bibfnamefont {R.}~\bibnamefont {Bassi}}, \ and\ \bibinfo {author} {\bibfnamefont {A.~R.}\ \bibnamefont {Holzwarth}},\ }\href {\doibase 10.1016/s0006-3495(01)76069-9} {\bibfield  {journal} {\bibinfo  {journal} {Biophys. J.}\ }\textbf {\bibinfo {volume} {80}},\ \bibinfo {pages} {901} (\bibinfo {year} {2001})}\BibitemShut {NoStop}%
\bibitem [{\citenamefont {Kaucikas}\ \emph {et~al.}(2017)\citenamefont {Kaucikas}, \citenamefont {Nurnberg}, \citenamefont {Dorlhiac}, \citenamefont {Rutherford},\ and\ \citenamefont {van Thor}}]{Kaucikas2017}%
  \BibitemOpen
  \bibfield  {author} {\bibinfo {author} {\bibfnamefont {M.}~\bibnamefont {Kaucikas}}, \bibinfo {author} {\bibfnamefont {D.}~\bibnamefont {Nurnberg}}, \bibinfo {author} {\bibfnamefont {G.}~\bibnamefont {Dorlhiac}}, \bibinfo {author} {\bibfnamefont {A.~W.}\ \bibnamefont {Rutherford}}, \ and\ \bibinfo {author} {\bibfnamefont {J.~J.}\ \bibnamefont {van Thor}},\ }\href {\doibase 10.1016/j.bpj.2016.12.022} {\bibfield  {journal} {\bibinfo  {journal} {Biophys. J.}\ }\textbf {\bibinfo {volume} {112}},\ \bibinfo {pages} {234} (\bibinfo {year} {2017})}\BibitemShut {NoStop}%
\bibitem [{\citenamefont {Zamzam}\ \emph {et~al.}(2020)\citenamefont {Zamzam}, \citenamefont {Rakowski}, \citenamefont {Kaucikas}, \citenamefont {Dorlhiac}, \citenamefont {Viola}, \citenamefont {Nurnberg}, \citenamefont {Fantuzzi}, \citenamefont {Rutherford},\ and\ \citenamefont {van Thor}}]{Zamzam2020}%
  \BibitemOpen
  \bibfield  {author} {\bibinfo {author} {\bibfnamefont {N.}~\bibnamefont {Zamzam}}, \bibinfo {author} {\bibfnamefont {R.}~\bibnamefont {Rakowski}}, \bibinfo {author} {\bibfnamefont {M.}~\bibnamefont {Kaucikas}}, \bibinfo {author} {\bibfnamefont {G.}~\bibnamefont {Dorlhiac}}, \bibinfo {author} {\bibfnamefont {S.}~\bibnamefont {Viola}}, \bibinfo {author} {\bibfnamefont {D.~J.}\ \bibnamefont {Nurnberg}}, \bibinfo {author} {\bibfnamefont {A.}~\bibnamefont {Fantuzzi}}, \bibinfo {author} {\bibfnamefont {A.~W.}\ \bibnamefont {Rutherford}}, \ and\ \bibinfo {author} {\bibfnamefont {J.~J.}\ \bibnamefont {van Thor}},\ }\href {\doibase 10.1073/pnas.2006016117} {\bibfield  {journal} {\bibinfo  {journal} {Proc. Natl. Acad. Sci.}\ }\textbf {\bibinfo {volume} {117}},\ \bibinfo {pages} {23158} (\bibinfo {year} {2020})}\BibitemShut {NoStop}%
\bibitem [{\citenamefont {Hunt}\ and\ \citenamefont {Dawlaty}(2019)}]{Hunt2019}%
  \BibitemOpen
  \bibfield  {author} {\bibinfo {author} {\bibfnamefont {J.~R.}\ \bibnamefont {Hunt}}\ and\ \bibinfo {author} {\bibfnamefont {J.~M.}\ \bibnamefont {Dawlaty}},\ }\href {\doibase 10.1021/acs.jpca.9b08970} {\bibfield  {journal} {\bibinfo  {journal} {J. Phys. Chem.}\ }\textbf {\bibinfo {volume} {123}},\ \bibinfo {pages} {10372} (\bibinfo {year} {2019})}\BibitemShut {NoStop}%
\bibitem [{\citenamefont {Jiang}\ \emph {et~al.}(2013)\citenamefont {Jiang}, \citenamefont {Liu}, \citenamefont {Song}, \citenamefont {He}, \citenamefont {Wang},\ and\ \citenamefont {Yang}}]{Jiang2013}%
  \BibitemOpen
  \bibfield  {author} {\bibinfo {author} {\bibfnamefont {L.}~\bibnamefont {Jiang}}, \bibinfo {author} {\bibfnamefont {W.}~\bibnamefont {Liu}}, \bibinfo {author} {\bibfnamefont {Y.}~\bibnamefont {Song}}, \bibinfo {author} {\bibfnamefont {X.}~\bibnamefont {He}}, \bibinfo {author} {\bibfnamefont {Y.}~\bibnamefont {Wang}}, \ and\ \bibinfo {author} {\bibfnamefont {Y.}~\bibnamefont {Yang}},\ }\href {\doibase 10.1007/s00340-013-5683-z} {\bibfield  {journal} {\bibinfo  {journal} {Appl. Phys. B}\ }\textbf {\bibinfo {volume} {116}},\ \bibinfo {pages} {271} (\bibinfo {year} {2013})}\BibitemShut {NoStop}%
\bibitem [{\citenamefont {Dobryakov}\ \emph {et~al.}(2005)\citenamefont {Dobryakov}, \citenamefont {Kovalenko},\ and\ \citenamefont {Ernsting}}]{Dobryakov2005}%
  \BibitemOpen
  \bibfield  {author} {\bibinfo {author} {\bibfnamefont {A.~L.}\ \bibnamefont {Dobryakov}}, \bibinfo {author} {\bibfnamefont {S.~A.}\ \bibnamefont {Kovalenko}}, \ and\ \bibinfo {author} {\bibfnamefont {N.~P.}\ \bibnamefont {Ernsting}},\ }\href {\doibase 10.1063/1.1948383} {\bibfield  {journal} {\bibinfo  {journal} {J. Chem. Phys.}\ }\textbf {\bibinfo {volume} {123}},\ \bibinfo {pages} {044502} (\bibinfo {year} {2005})}\BibitemShut {NoStop}%
\bibitem [{\citenamefont {Maly}\ \emph {et~al.}(2017)\citenamefont {Maly}, \citenamefont {Ravensbergen}, \citenamefont {Kennis}, \citenamefont {van Grondelle}, \citenamefont {Croce}, \citenamefont {Mancal},\ and\ \citenamefont {van Oort}}]{Maly2017}%
  \BibitemOpen
  \bibfield  {author} {\bibinfo {author} {\bibfnamefont {P.}~\bibnamefont {Maly}}, \bibinfo {author} {\bibfnamefont {J.}~\bibnamefont {Ravensbergen}}, \bibinfo {author} {\bibfnamefont {J.~T.~M.}\ \bibnamefont {Kennis}}, \bibinfo {author} {\bibfnamefont {R.}~\bibnamefont {van Grondelle}}, \bibinfo {author} {\bibfnamefont {R.}~\bibnamefont {Croce}}, \bibinfo {author} {\bibfnamefont {T.}~\bibnamefont {Mancal}}, \ and\ \bibinfo {author} {\bibfnamefont {B.}~\bibnamefont {van Oort}},\ }\href {\doibase 10.1038/srep43484} {\bibfield  {journal} {\bibinfo  {journal} {Sci. Rep.}\ }\textbf {\bibinfo {volume} {7}},\ \bibinfo {pages} {43484} (\bibinfo {year} {2017})}\BibitemShut {NoStop}%
\bibitem [{\citenamefont {Ohkita}\ and\ \citenamefont {Ito}(2011)}]{Ohkita2011}%
  \BibitemOpen
  \bibfield  {author} {\bibinfo {author} {\bibfnamefont {H.}~\bibnamefont {Ohkita}}\ and\ \bibinfo {author} {\bibfnamefont {S.}~\bibnamefont {Ito}},\ }\href {\doibase 10.1016/j.polymer.2011.06.061} {\bibfield  {journal} {\bibinfo  {journal} {Polymer}\ }\textbf {\bibinfo {volume} {52}},\ \bibinfo {pages} {4397} (\bibinfo {year} {2011})}\BibitemShut {NoStop}%
\bibitem [{\citenamefont {Ravensbergen}\ \emph {et~al.}(2014)\citenamefont {Ravensbergen}, \citenamefont {Abdi}, \citenamefont {van Santen}, \citenamefont {Frese}, \citenamefont {Dam}, \citenamefont {van~de Krol},\ and\ \citenamefont {Kennis}}]{Ravensbergen2014}%
  \BibitemOpen
  \bibfield  {author} {\bibinfo {author} {\bibfnamefont {J.}~\bibnamefont {Ravensbergen}}, \bibinfo {author} {\bibfnamefont {F.~F.}\ \bibnamefont {Abdi}}, \bibinfo {author} {\bibfnamefont {J.~H.}\ \bibnamefont {van Santen}}, \bibinfo {author} {\bibfnamefont {R.~N.}\ \bibnamefont {Frese}}, \bibinfo {author} {\bibfnamefont {B.}~\bibnamefont {Dam}}, \bibinfo {author} {\bibfnamefont {R.}~\bibnamefont {van~de Krol}}, \ and\ \bibinfo {author} {\bibfnamefont {J.~T.~M.}\ \bibnamefont {Kennis}},\ }\href {\doibase 10.1021/jp509930s} {\bibfield  {journal} {\bibinfo  {journal} {J. Phys. Chem. C}\ }\textbf {\bibinfo {volume} {118}},\ \bibinfo {pages} {27793} (\bibinfo {year} {2014})}\BibitemShut {NoStop}%
\bibitem [{\citenamefont {Knowles}\ \emph {et~al.}(2018)\citenamefont {Knowles}, \citenamefont {Koch},\ and\ \citenamefont {Shelton}}]{Knowles2018}%
  \BibitemOpen
  \bibfield  {author} {\bibinfo {author} {\bibfnamefont {K.~E.}\ \bibnamefont {Knowles}}, \bibinfo {author} {\bibfnamefont {M.~D.}\ \bibnamefont {Koch}}, \ and\ \bibinfo {author} {\bibfnamefont {J.~L.}\ \bibnamefont {Shelton}},\ }\href {\doibase 10.1039/c8tc02977f} {\bibfield  {journal} {\bibinfo  {journal} {J. Mater. Chem. C}\ }\textbf {\bibinfo {volume} {6}},\ \bibinfo {pages} {11853} (\bibinfo {year} {2018})}\BibitemShut {NoStop}%
\bibitem [{\citenamefont {Yarita}\ \emph {et~al.}(2017)\citenamefont {Yarita}, \citenamefont {Tahara}, \citenamefont {Ihara}, \citenamefont {Kawawaki}, \citenamefont {Sato}, \citenamefont {Saruyama}, \citenamefont {Teranishi},\ and\ \citenamefont {Kanemitsu}}]{Yarita2017}%
  \BibitemOpen
  \bibfield  {author} {\bibinfo {author} {\bibfnamefont {N.}~\bibnamefont {Yarita}}, \bibinfo {author} {\bibfnamefont {H.}~\bibnamefont {Tahara}}, \bibinfo {author} {\bibfnamefont {T.}~\bibnamefont {Ihara}}, \bibinfo {author} {\bibfnamefont {T.}~\bibnamefont {Kawawaki}}, \bibinfo {author} {\bibfnamefont {R.}~\bibnamefont {Sato}}, \bibinfo {author} {\bibfnamefont {M.}~\bibnamefont {Saruyama}}, \bibinfo {author} {\bibfnamefont {T.}~\bibnamefont {Teranishi}}, \ and\ \bibinfo {author} {\bibfnamefont {Y.}~\bibnamefont {Kanemitsu}},\ }\href {\doibase 10.1021/acs.jpclett.7b00326} {\bibfield  {journal} {\bibinfo  {journal} {J. Phys. Chem. Lett.}\ }\textbf {\bibinfo {volume} {8}},\ \bibinfo {pages} {1413} (\bibinfo {year} {2017})}\BibitemShut {NoStop}%
\bibitem [{\citenamefont {Wang}\ \emph {et~al.}(2014)\citenamefont {Wang}, \citenamefont {McCleese}, \citenamefont {Kovalsky}, \citenamefont {Zhao},\ and\ \citenamefont {Burda}}]{Wang2014}%
  \BibitemOpen
  \bibfield  {author} {\bibinfo {author} {\bibfnamefont {L.}~\bibnamefont {Wang}}, \bibinfo {author} {\bibfnamefont {C.}~\bibnamefont {McCleese}}, \bibinfo {author} {\bibfnamefont {A.}~\bibnamefont {Kovalsky}}, \bibinfo {author} {\bibfnamefont {Y.}~\bibnamefont {Zhao}}, \ and\ \bibinfo {author} {\bibfnamefont {C.}~\bibnamefont {Burda}},\ }\href {\doibase 10.1021/ja504632z} {\bibfield  {journal} {\bibinfo  {journal} {J. Am. Chem. Soc.}\ }\textbf {\bibinfo {volume} {136}},\ \bibinfo {pages} {12205} (\bibinfo {year} {2014})}\BibitemShut {NoStop}%
\bibitem [{\citenamefont {Serpetzoglou}\ \emph {et~al.}(2017)\citenamefont {Serpetzoglou}, \citenamefont {Konidakis}, \citenamefont {Kakavelakis}, \citenamefont {Maksudov}, \citenamefont {Kymakis},\ and\ \citenamefont {Stratakis}}]{Serpetzoglou2017}%
  \BibitemOpen
  \bibfield  {author} {\bibinfo {author} {\bibfnamefont {E.}~\bibnamefont {Serpetzoglou}}, \bibinfo {author} {\bibfnamefont {I.}~\bibnamefont {Konidakis}}, \bibinfo {author} {\bibfnamefont {G.}~\bibnamefont {Kakavelakis}}, \bibinfo {author} {\bibfnamefont {T.}~\bibnamefont {Maksudov}}, \bibinfo {author} {\bibfnamefont {E.}~\bibnamefont {Kymakis}}, \ and\ \bibinfo {author} {\bibfnamefont {E.}~\bibnamefont {Stratakis}},\ }\href {\doibase 10.1021/acsami.7b15195} {\bibfield  {journal} {\bibinfo  {journal} {Acs. Appl. Mater. Inter.}\ }\textbf {\bibinfo {volume} {9}},\ \bibinfo {pages} {43910} (\bibinfo {year} {2017})}\BibitemShut {NoStop}%
\bibitem [{\citenamefont {Bridgman}(1950)}]{Bridgman1950}%
  \BibitemOpen
  \bibfield  {author} {\bibinfo {author} {\bibfnamefont {P.~W.}\ \bibnamefont {Bridgman}},\ }\href@noop {} {\bibfield  {journal} {\bibinfo  {journal} {P. Roy. Soc. A-math. Phy.}\ }\textbf {\bibinfo {volume} {203}},\ \bibinfo {pages} {1} (\bibinfo {year} {1950})}\BibitemShut {NoStop}%
\bibitem [{\citenamefont {Bassett}(2009)}]{Bassett2009}%
  \BibitemOpen
  \bibfield  {author} {\bibinfo {author} {\bibfnamefont {W.~A.}\ \bibnamefont {Bassett}},\ }\href {\doibase 10.1080/08957950802597239} {\bibfield  {journal} {\bibinfo  {journal} {High Pressure Res.}\ }\textbf {\bibinfo {volume} {29}},\ \bibinfo {pages} {163} (\bibinfo {year} {2009})}\BibitemShut {NoStop}%
\bibitem [{\citenamefont {Jayaraman}(1983)}]{Jayaraman1983}%
  \BibitemOpen
  \bibfield  {author} {\bibinfo {author} {\bibfnamefont {A.}~\bibnamefont {Jayaraman}},\ }\href {\doibase 10.1103/revmodphys.55.65} {\bibfield  {journal} {\bibinfo  {journal} {Rev. Mod. Phys}\ }\textbf {\bibinfo {volume} {55}},\ \bibinfo {pages} {65} (\bibinfo {year} {1983})}\BibitemShut {NoStop}%
\bibitem [{\citenamefont {Zhou}\ \emph {et~al.}(2020)\citenamefont {Zhou}, \citenamefont {Cao}, \citenamefont {Li}, \citenamefont {Li}, \citenamefont {Yin}, \citenamefont {Gao}, \citenamefont {Jin}, \citenamefont {Shi}, \citenamefont {Liu},\ and\ \citenamefont {Ding}}]{Zhou2020}%
  \BibitemOpen
  \bibfield  {author} {\bibinfo {author} {\bibfnamefont {Q.}~\bibnamefont {Zhou}}, \bibinfo {author} {\bibfnamefont {B.}~\bibnamefont {Cao}}, \bibinfo {author} {\bibfnamefont {Y.}~\bibnamefont {Li}}, \bibinfo {author} {\bibfnamefont {B.}~\bibnamefont {Li}}, \bibinfo {author} {\bibfnamefont {H.}~\bibnamefont {Yin}}, \bibinfo {author} {\bibfnamefont {J.}~\bibnamefont {Gao}}, \bibinfo {author} {\bibfnamefont {M.}~\bibnamefont {Jin}}, \bibinfo {author} {\bibfnamefont {Y.}~\bibnamefont {Shi}}, \bibinfo {author} {\bibfnamefont {C.}~\bibnamefont {Liu}}, \ and\ \bibinfo {author} {\bibfnamefont {D.}~\bibnamefont {Ding}},\ }\href {\doibase 10.1063/5.0015843} {\bibfield  {journal} {\bibinfo  {journal} {Appl. Phys. Lett.}\ }\textbf {\bibinfo {volume} {117}},\ \bibinfo {pages} {063101} (\bibinfo {year} {2020})}\BibitemShut {NoStop}%
\bibitem [{\citenamefont {Liu}\ \emph {et~al.}(2023)\citenamefont {Liu}, \citenamefont {Dong}, \citenamefont {Zhang}, \citenamefont {He}, \citenamefont {Li}, \citenamefont {Niu}, \citenamefont {Yu},\ and\ \citenamefont {Jia}}]{Liu2023}%
  \BibitemOpen
  \bibfield  {author} {\bibinfo {author} {\bibfnamefont {D.}~\bibnamefont {Liu}}, \bibinfo {author} {\bibfnamefont {D.}~\bibnamefont {Dong}}, \bibinfo {author} {\bibfnamefont {Z.}~\bibnamefont {Zhang}}, \bibinfo {author} {\bibfnamefont {Y.}~\bibnamefont {He}}, \bibinfo {author} {\bibfnamefont {Z.}~\bibnamefont {Li}}, \bibinfo {author} {\bibfnamefont {J.}~\bibnamefont {Niu}}, \bibinfo {author} {\bibfnamefont {N.}~\bibnamefont {Yu}}, \ and\ \bibinfo {author} {\bibfnamefont {H.}~\bibnamefont {Jia}},\ }\href {\doibase 10.1016/j.optmat.2023.114537} {\bibfield  {journal} {\bibinfo  {journal} {Opt. Mater.}\ }\textbf {\bibinfo {volume} {146}},\ \bibinfo {pages} {114537} (\bibinfo {year} {2023})}\BibitemShut {NoStop}%
\bibitem [{\citenamefont {Wu}\ \emph {et~al.}(2021)\citenamefont {Wu}, \citenamefont {Yin}, \citenamefont {Hasaien}, \citenamefont {Tian}, \citenamefont {Ding},\ and\ \citenamefont {Zhao}}]{Wu2021}%
  \BibitemOpen
  \bibfield  {author} {\bibinfo {author} {\bibfnamefont {Y.~L.}\ \bibnamefont {Wu}}, \bibinfo {author} {\bibfnamefont {X.}~\bibnamefont {Yin}}, \bibinfo {author} {\bibfnamefont {J.~Z.~L.}\ \bibnamefont {Hasaien}}, \bibinfo {author} {\bibfnamefont {Z.~Y.}\ \bibnamefont {Tian}}, \bibinfo {author} {\bibfnamefont {Y.}~\bibnamefont {Ding}}, \ and\ \bibinfo {author} {\bibfnamefont {J.}~\bibnamefont {Zhao}},\ }\href {\doibase 10.1063/5.0064071} {\bibfield  {journal} {\bibinfo  {journal} {Rev. Sci. Instrum.}\ }\textbf {\bibinfo {volume} {92}},\ \bibinfo {pages} {113002} (\bibinfo {year} {2021})}\BibitemShut {NoStop}%
\bibitem [{\citenamefont {Li}\ \emph {et~al.}(2020)\citenamefont {Li}, \citenamefont {Sui}, \citenamefont {Niu}, \citenamefont {Jiang}, \citenamefont {Zhang}, \citenamefont {Wu}, \citenamefont {Jin},\ and\ \citenamefont {Yuan}}]{Li2020}%
  \BibitemOpen
  \bibfield  {author} {\bibinfo {author} {\bibfnamefont {Q.}~\bibnamefont {Li}}, \bibinfo {author} {\bibfnamefont {L.}~\bibnamefont {Sui}}, \bibinfo {author} {\bibfnamefont {G.}~\bibnamefont {Niu}}, \bibinfo {author} {\bibfnamefont {J.}~\bibnamefont {Jiang}}, \bibinfo {author} {\bibfnamefont {Y.}~\bibnamefont {Zhang}}, \bibinfo {author} {\bibfnamefont {G.}~\bibnamefont {Wu}}, \bibinfo {author} {\bibfnamefont {M.}~\bibnamefont {Jin}}, \ and\ \bibinfo {author} {\bibfnamefont {K.}~\bibnamefont {Yuan}},\ }\href {\doibase 10.1021/acs.jpcc.0c01869} {\bibfield  {journal} {\bibinfo  {journal} {J. Phys. Chem. C}\ }\textbf {\bibinfo {volume} {124}},\ \bibinfo {pages} {11183} (\bibinfo {year} {2020})}\BibitemShut {NoStop}%
\bibitem [{\citenamefont {Dubietis}\ and\ \citenamefont {Couairon}(2019)}]{Dubietis2019}%
  \BibitemOpen
  \bibfield  {author} {\bibinfo {author} {\bibfnamefont {A.}~\bibnamefont {Dubietis}}\ and\ \bibinfo {author} {\bibfnamefont {A.}~\bibnamefont {Couairon}},\ }\href@noop {} {\emph {\bibinfo {title} {Ultrafast Supercontinuum Generation in Transparent Solid-State Media}}},\ SpringerBriefs in Physics\ (\bibinfo  {publisher} {Springer International Publishing},\ \bibinfo {year} {2019})\BibitemShut {NoStop}%
\bibitem [{\citenamefont {Dharmadhikari}\ \emph {et~al.}(2005)\citenamefont {Dharmadhikari}, \citenamefont {Rajgara},\ and\ \citenamefont {Mathur}}]{Dharmadhikari2005}%
  \BibitemOpen
  \bibfield  {author} {\bibinfo {author} {\bibfnamefont {A.}~\bibnamefont {Dharmadhikari}}, \bibinfo {author} {\bibfnamefont {F.}~\bibnamefont {Rajgara}}, \ and\ \bibinfo {author} {\bibfnamefont {D.}~\bibnamefont {Mathur}},\ }\href {\doibase 10.1007/s00340-004-1682-4} {\bibfield  {journal} {\bibinfo  {journal} {Appl. Phys. B}\ }\textbf {\bibinfo {volume} {80}},\ \bibinfo {pages} {61} (\bibinfo {year} {2005})}\BibitemShut {NoStop}%
\bibitem [{\citenamefont {Schott}\ \emph {et~al.}(2014)\citenamefont {Schott}, \citenamefont {Steinbacher}, \citenamefont {Buback}, \citenamefont {Nuernberger},\ and\ \citenamefont {Brixner}}]{Schott2014}%
  \BibitemOpen
  \bibfield  {author} {\bibinfo {author} {\bibfnamefont {S.}~\bibnamefont {Schott}}, \bibinfo {author} {\bibfnamefont {A.}~\bibnamefont {Steinbacher}}, \bibinfo {author} {\bibfnamefont {J.}~\bibnamefont {Buback}}, \bibinfo {author} {\bibfnamefont {P.}~\bibnamefont {Nuernberger}}, \ and\ \bibinfo {author} {\bibfnamefont {T.}~\bibnamefont {Brixner}},\ }\href {\doibase 10.1088/0953-4075/47/12/124014} {\bibfield  {journal} {\bibinfo  {journal} {J. Phys. B}\ }\textbf {\bibinfo {volume} {47}},\ \bibinfo {pages} {124014} (\bibinfo {year} {2014})}\BibitemShut {NoStop}%
\bibitem [{\citenamefont {Tobias}\ \emph {et~al.}(2004)\citenamefont {Tobias}, \citenamefont {Tom\'a\v{s}}, \citenamefont {Stiopkinand},\ and\ \citenamefont {Fleming}}]{Brixner2004}%
  \BibitemOpen
  \bibfield  {author} {\bibinfo {author} {\bibfnamefont {B.}~\bibnamefont {Tobias}}, \bibinfo {author} {\bibfnamefont {M.}~\bibnamefont {Tom\'a\v{s}}}, \bibinfo {author} {\bibfnamefont {I.~V.}\ \bibnamefont {Stiopkinand}}, \ and\ \bibinfo {author} {\bibfnamefont {G.~R.}\ \bibnamefont {Fleming}},\ }\href {\doibase 10.1063/1.1776112} {\bibfield  {journal} {\bibinfo  {journal} {J. Chem. Phys.}\ }\textbf {\bibinfo {volume} {121}},\ \bibinfo {pages} {4221} (\bibinfo {year} {2004})}\BibitemShut {NoStop}%
\bibitem [{\citenamefont {Yue}\ \emph {et~al.}(2022)\citenamefont {Yue}, \citenamefont {Zhou}, \citenamefont {Su},\ and\ \citenamefont {Zhang}}]{Yue2022}%
  \BibitemOpen
  \bibfield  {author} {\bibinfo {author} {\bibfnamefont {J.}~\bibnamefont {Yue}}, \bibinfo {author} {\bibfnamefont {L.}~\bibnamefont {Zhou}}, \bibinfo {author} {\bibfnamefont {P.}~\bibnamefont {Su}}, \ and\ \bibinfo {author} {\bibfnamefont {W.}~\bibnamefont {Zhang}},\ }\href {\doibase 10.1016/j.cplett.2022.139766} {\bibfield  {journal} {\bibinfo  {journal} {Chem. Phys. Lett.}\ }\textbf {\bibinfo {volume} {802}},\ \bibinfo {pages} {139766} (\bibinfo {year} {2022})}\BibitemShut {NoStop}%
\bibitem [{\citenamefont {Fujii}\ \emph {et~al.}(1995)\citenamefont {Fujii}, \citenamefont {Nishikiori},\ and\ \citenamefont {Tamura}}]{Fujii1995}%
  \BibitemOpen
  \bibfield  {author} {\bibinfo {author} {\bibfnamefont {T.}~\bibnamefont {Fujii}}, \bibinfo {author} {\bibfnamefont {H.}~\bibnamefont {Nishikiori}}, \ and\ \bibinfo {author} {\bibfnamefont {T.}~\bibnamefont {Tamura}},\ }\href {\doibase 10.1016/0009-2614(94)01477-d} {\bibfield  {journal} {\bibinfo  {journal} {Chem. Phys. Lett.}\ }\textbf {\bibinfo {volume} {233}},\ \bibinfo {pages} {424} (\bibinfo {year} {1995})}\BibitemShut {NoStop}%
\bibitem [{\citenamefont {Roberts}\ and\ \citenamefont {Drickamer}(1985)}]{Roberts1985}%
  \BibitemOpen
  \bibfield  {author} {\bibinfo {author} {\bibfnamefont {E.~R.}\ \bibnamefont {Roberts}}\ and\ \bibinfo {author} {\bibfnamefont {H.~G.}\ \bibnamefont {Drickamer}},\ }\href {\doibase 10.1021/j100260a028} {\bibfield  {journal} {\bibinfo  {journal} {J. Phys. Chem.}\ }\textbf {\bibinfo {volume} {89}},\ \bibinfo {pages} {3092} (\bibinfo {year} {1985})}\BibitemShut {NoStop}%
\bibitem [{\citenamefont {Kristoffersen}\ \emph {et~al.}(2014)\citenamefont {Kristoffersen}, \citenamefont {Erga}, \citenamefont {Hamre},\ and\ \citenamefont {Frette}}]{Kristoffersen2014}%
  \BibitemOpen
  \bibfield  {author} {\bibinfo {author} {\bibfnamefont {A.~S.}\ \bibnamefont {Kristoffersen}}, \bibinfo {author} {\bibfnamefont {S.~R.}\ \bibnamefont {Erga}}, \bibinfo {author} {\bibfnamefont {B.}~\bibnamefont {Hamre}}, \ and\ \bibinfo {author} {\bibfnamefont {{\O}.}~\bibnamefont {Frette}},\ }\href {\doibase 10.1007/s10895-014-1368-1} {\bibfield  {journal} {\bibinfo  {journal} {J. Fluoresc.}\ }\textbf {\bibinfo {volume} {24}},\ \bibinfo {pages} {1015} (\bibinfo {year} {2014})}\BibitemShut {NoStop}%
\bibitem [{\citenamefont {del Monte}\ and\ \citenamefont {Levy}(1998)}]{Monte1998}%
  \BibitemOpen
  \bibfield  {author} {\bibinfo {author} {\bibfnamefont {F.}~\bibnamefont {del Monte}}\ and\ \bibinfo {author} {\bibfnamefont {D.}~\bibnamefont {Levy}},\ }\href {\doibase 10.1021/jp982396v} {\bibfield  {journal} {\bibinfo  {journal} {J. Phys. Chem. B}\ }\textbf {\bibinfo {volume} {102}},\ \bibinfo {pages} {8036} (\bibinfo {year} {1998})}\BibitemShut {NoStop}%
\end{thebibliography}%

\end{document}